# Design and Implementation of Wireless Energy Meter System for Monitoring the Single Phase Supply

Prashanth B.U.V
Instrumentation Engineer
Dept. of Science & Technology (DST)
Promotion of University Research & Scientific Excellence (PURSE)
Sri Venkateswara University, Tirupati-517502.

## ABSTRACT
Wireless energy meter is a system developed to serve as a basic single-phase energy meter with advanced functionalities such as Peak hour setting, Peak load setting Wireless reading transmission; further the system eliminates the role of a Meter Reader.

## 1. INTRODUCTION
In the present scenario, there is no option for the electricity board to regulate the load distribution during peak hours. As a result, even today, a number of customers are suffering due to Low Voltage during peak hours. Power is saved at the mercy of customers only during peak hours. A large amount is spend for obtaining the meter readings every month. Meter readers find it difficult to collect the readings from remote areas. The wireless energy meter product specification is the energy meter with increased functionality of a single phase handheld module monitoring system.

## 2. PRINCIPLE
WEM monitors the energy consumption for a single phase supply. There will be a preset value for the Peak hour & permissible load at peak hour. Bimonthly readings will be sent to the control station on the second day [1]. At the base station, the total bill amount is prepared. Further at the base station, NCU & ECU are extracted. The energy meter module consists of the basic functions of standard energy meter, It consists of a Single phase supply, It is having the capability of Unit & Time setting ,Further it has a detachable Keypad facility along with GSM facility and LCD display[1].

## 3. DESIGN OF WEM

In this design of wireless energy meter the ADE7752 used as energy metering IC. Inputs to the IC are the stepped down voltage & current of the utility supply using CT & PT. The IC generates pulses as output proportional to power consumed. Micro reads the number of pulses as 3200=1 unit of energy. The circuit shown in figure1 is the circuit of energy metering IC ADE7752. Below circuit shows the power supply circuit using the regulator IC LM7805.

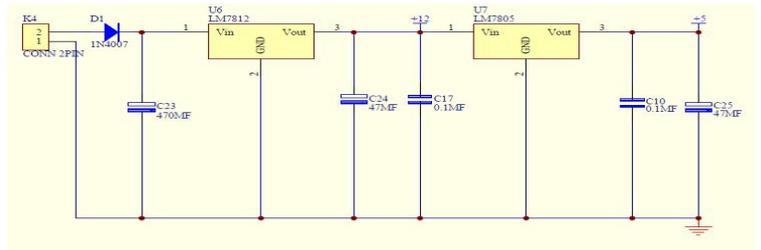

**Figure 2 Power Supply Circuit.**

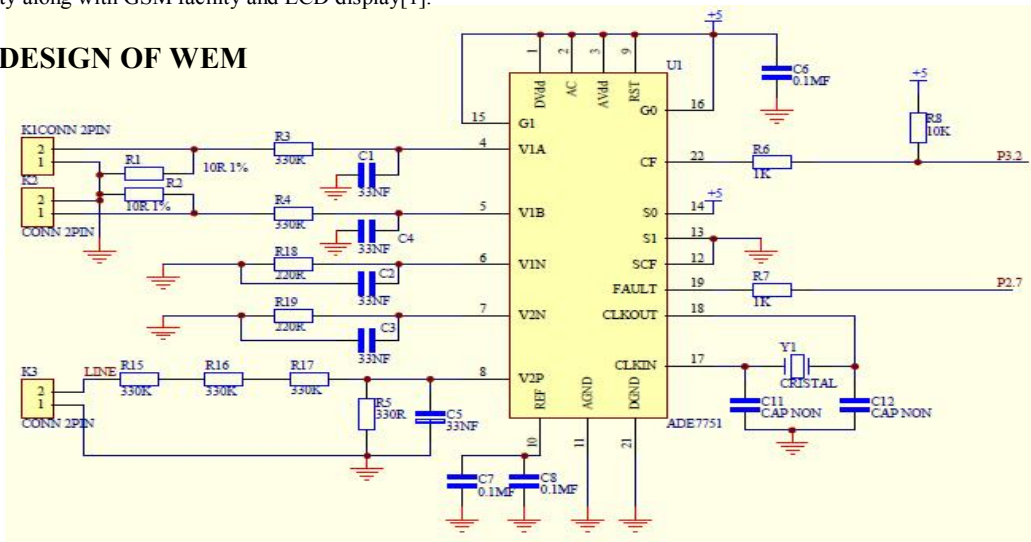

**Figure 1. Circuit Diagram of energy metering IC ADE7751**

As shown in the figure3 the microcontroller circuit is interfaced to a 16x2 LCD and a resistor pack connected to Vcc of +5V. Further a MAX232 IC is interfaced to the microcontroller for the interface to the serial to the P.C with baud rate of 9600 bps. This circuit is as shown in the next figure4.Here the MAX232 is interfaced to the port 3 of the microcontroller AT89C52.

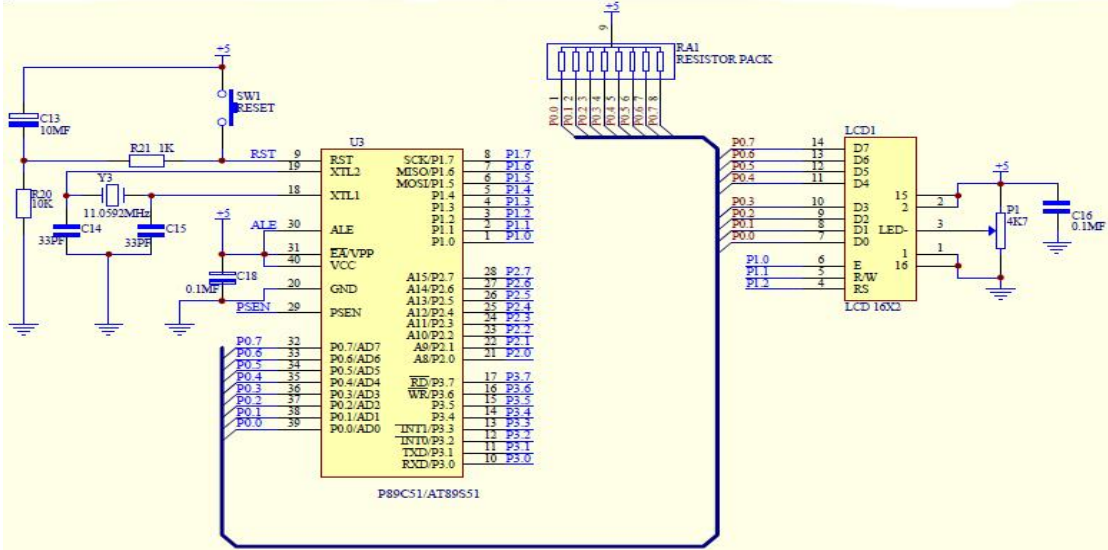

**Figure 3 Microcontroller Circuit**

Further the MAX232 circuit is used to interface with the external devices such as P.C with hyper terminal software as the interfacing software with the baud rate of 9600kbps [5].This circuit is as shown in figure 4.

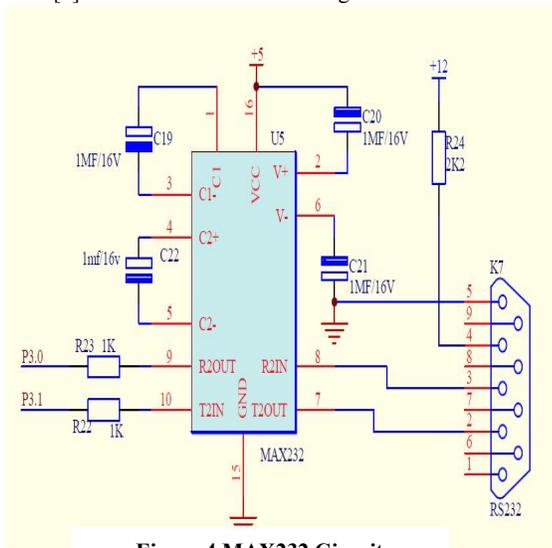

**Figure 4 MAX232 Circuit**

Further the entire module is interfaced to the real time clock DS1307 as shown in the below figure.

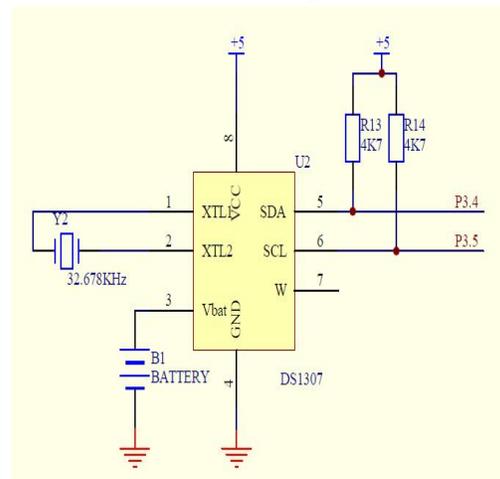

**Figure 5 Real Time Clock circuit**

Energy meter Consists of Atmel micro controller which has a long period life cycle. RTC IC (DS 1307) connected with a 3.3V Battery which will last for a minimum period of 10 years unless any short circuit occurs. So the circuit is more reliable.

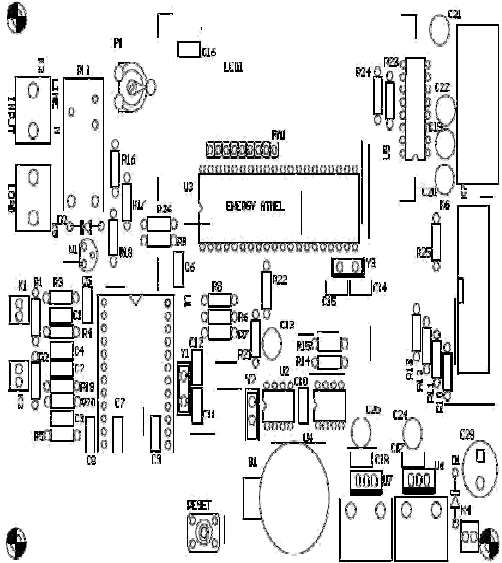

**Figure 6 PCB Module of the entire circuit.**

## 4. COMMUNICATION PROTOCOL.

Wireless communication with the control station is made possible by means of the GSM modem. It supports the AT command sets. Using Serial communication protocol (USART) we communicate with the GSM Modem. RTC and EEPROM [8]. The meter reading are stored in each change in values to a permanent memory. This is made possible by means of $I^2C$ protocol. It's a 3 wire synchronous communication protocol. A scanning method is used to interface with the keyboard.

Steps to Interface with MODEM

Step 1:-GSM modem supports the AT command set AT + (0x0d, 0x0a) is send to the modem. Modem returned with an OK signal in the same baud rate.

Step 2: Send ATE0 + (0x0d, 0x0a) to the modem to Turn the Echo off Modem returned with an OK signal in the same baud rate.

Step3: Send command AT+CMGF=1 + (0x0d, 0x0a) to Select text mode in the GSM Modem returned with an OK signal in the same baud rate.

Step 4: To send the Message use the following format

AT+CMGS="+919xxxxxxxxx", after getting a confirmation (on receiving > from modem) s end the Meter reading and meter ID.

## 5. ADVANTAGES OF WEM

Even though a number of Energy meters are available, WEM has following unique features such as Peak Hour-Load setting capability, further it completely avoids the role of meter reader, also it has the features of Wireless Data logging with previous data base search option.

## 6. OPERATION OF WEM

Energy Meter tested at different voltage conditions using an Auto transformer. Supply voltage varied from 150 VAC to 240 VAC. During this testing process meter consistently works with accuracy. Different load devices are connected to test the current handling capability of the meter (10W to 1000W).

## 7. RESULTS

The hardware design of the energy meter has been completed and implemented. The code at meter side completed and tested. The front end is designed using Visual basic [2]. Mobile phones are used instead of GSM modules. The test results using Visual basic screen shots are as follows.

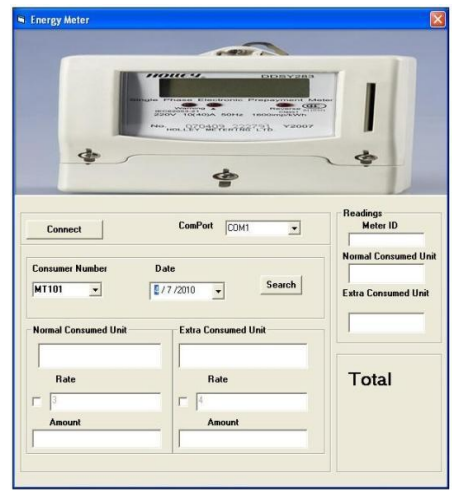

**Figure 7 software front end**

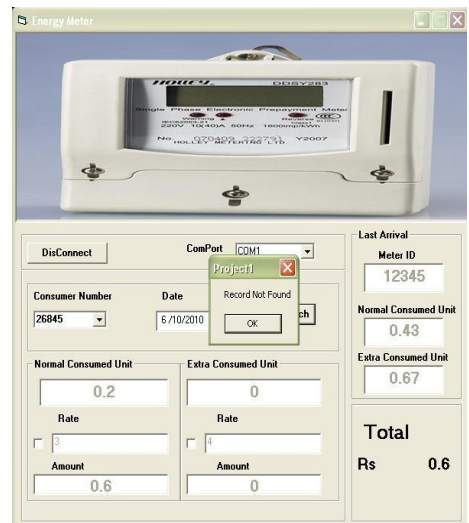

**Figure 8 Invalid Entry screen.**

E. Once the pass word is entered
  There will be options to edit through keypad
  That is as follows
a. ID
b. Fixed unit value
c. Mobile Number
d. Exit

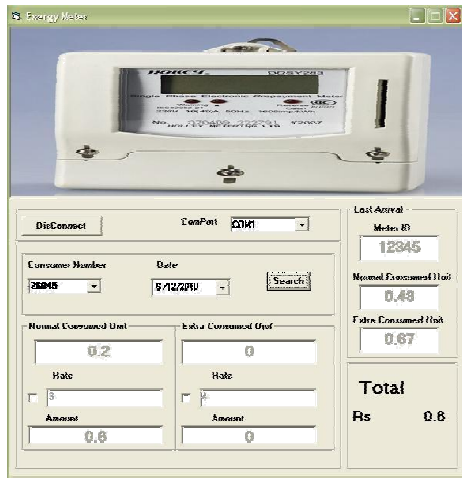

**Figure9 Front End Without extra unit.**

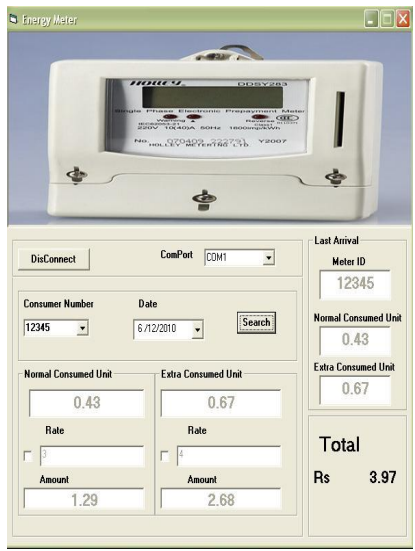

**Figure 10 Front End with extra unit.**

The operating conditions of WEM are as follows, the energy meter tested at different voltage conditions using an Auto transformer. Supply voltage varied from 150 VAC to 240 VAC. During this testing process meter consistently works with accuracy. Different load devices are connected to test the current handling capability of the meter (10W to 1000W).

The output results can be described as follows
  A. When wireless energy meter is switched ON, the "minute" value is set to 1.
     Initializations take place
  B. When "minute"=2, a message is sent, hourly.
     For each hour, 5-8 minutes taken as peak hours.
  C. Total reading displayed in the LCD and gets updated.
#$12345$00.00$00.00$* is the format for transmission ie;#$ meter id $ (total-extra) $ (extra) $*
  D. Default password is 1234

## 8. CONCLUSIONS

The past few decades have been marked by the onslaught of devices that are intelligent and interactive. The present day scenario the embedded systems are designed as hardware-software co-design, that is software is design is ported on the standalone board that contains a microcontroller with the other peripheral devices interfaced to it. The other peripheral devices interfaced are Real Time Clock (RTC), MAX232 circuit, Power supply circuit, along with LCD and other glue logic circuit. And finally taking all the above points into consideration a wireless energy meter (WEM) system is developed to serve as a basic single-phase energy meter. This meter is equipped with the advanced functionalities such as Peak hour setting, Peak load setting Wireless reading transmission. Further the system eliminates the role of a Meter Reader.

## 9. FUTURE SCOPE

The wireless energy meter system can modified to a 3-phase supply. In this system the printing mechanism can be embedded at the meter side so that a hardcopy of the bill may be obtained. A GSM module can be embedded in the same PCB as different layers [6] .The overall size can be reduced if multilayer PCB is used.